\documentclass[preprint,aps,superscriptaddress,nofootinbib]{revtex4-1}
\usepackage{graphicx}
\usepackage{dcolumn}   
\usepackage{bm}       
\usepackage{amssymb}
\usepackage{amsmath}
\usepackage[toc,page]{appendix}
\usepackage{braket}
\newcommand{\del}{\partial}

\begin{document}

\title{Taming Dirac strings and timelike loops in vacuum gravity
}

\author{Suvikranth Gera}
\email{suvikranthg@gmail.com}
\affiliation{Department of Physics, Indian Institute of Technology Kharagpur, Kharagpur-721302, INDIA }

\author{Sandipan Sengupta}
\email{sandipan@phy.iitkgp.ac.in}
\affiliation{Department of Physics and Centre for Theoretical Studies, Indian Institute of Technology Kharagpur, Kharagpur-721302, INDIA }

\begin{abstract}

The problem of singularities associated with Dirac strings and closed timelike curves in classical solutions of pure gravity is analyzed here. A method to eliminate these is introduced and established first for the Taub-NUT geometry. This is superceded by a smooth solution of first order field equations, which is defined to be a unique extension of the Taub Universe to a degenerate metric phase.
As an additional feature, this framework naturally provides a geometric interpretation of the magnetic charge in the context of gravity theory without matter. 
Finally, exploiting the two phases of the metric determinant, we find a (smooth and unique) continuation of  the Misner geometry as well, ridding it of  closed timelike worldlines which exist in its otherwise Einsteinian manifestation.

%


\end{abstract}

\maketitle

\section{Introduction}
Among the selected few exact vacuum solutions to Einsteinian gravity that defy a completely realistic physical interpretation, the Taub-NUT spacetime stands out in its richness and peculiarity \cite{taub,misner,misner1,nut,taub1,dowker,lynden}. Even though this geometry does not exhibit any divergence in the curvature polynomials, it carries within a Dirac string singularity, analogous to the magnetic monopole configuration in gauge theories. In addition, it is associated with closed timelike curves at the NUT region(s). As noted by Misner \cite{misner}, the string may be
removed by using a union of two overlapping charts, but only at the cost of introducing a periodic time coordinate. Not only does this introduce causality violating worldlines throughout the NUT region, but also leads to a nontrivial topology ($ S^3 \otimes R$) of the spacetime. There have been alternative viewpoints  as well, based on attempts to interpret the Taub-NUT geometry as a solution to Einstein's equation with a point mass and a massless source of angular momentum \cite{bonnor}. This interpretation corresponds to a spacetime different than that of Misner's, since the pathological regions containing the semi-infinite string and timelike loops are excluded from the manifold. Progress have also been made in unravelling several other key aspects, for instance, the possible (non-unique) maximal extensions of the Taub-NUT space \cite{taub1,miller} and the behaviour of its geodesics \cite{misner,misner1,zimmerman,kunz}. Nevertheless, it has not ceased to stoke up serious interest and intrigue till now, simply because it continues to be a challenge to find a satisfactory interpretation of this spacetime without having to live with a Dirac string or a direct violation of causality over a large region of spacetime \cite{misner1}.

With the purpose of providing a fresh perspective onto these issues, here we would like to explore whether it is possible to view the Taub-NUT geometry as a special phase within a more general spacetime. The only formulation of vacuum gravity where this could be possible is the first order theory, which, in addition to the Einsteinian (invertible metric) phase, admits a zero-determinant metric phase (with or without matter) \cite{tseytlin}. In view of this, we shall look to construct spacetime solutions to the field equations where the Taub phase could coexist with a noninvertible phase without ever evolving into a NUT geometry. We shall also analyze if such a special solution could be unique in any precise sense, such that it does not exhibit either a Dirac (Misner) string or any closed timelike curve. It is also essential that the basic gauge-invariant field components be at least continuous over the whole spacetime, so that the full geometry satisfies the equations of motion everywhere. Moreover, the field-strength components must be regular everywhere in order to ensure that there is no curvature singularity.

General solutions of Hilbert-Palatini field equations containing degenerate as well as Taub or NUT phases  have not been explored in the earlier literature. However, degenerate geometries based on stationary black hole solutions of Einstein's theory have been analyzed in the context of (complex) Sen-Ashtekar Hamiltonian formulation. Typically, these are associated with one or more zero eigenvalues of the canonical momenta (densitized triad $E^a_i$). For instance, Bengtsson have constructed a solution of the canonical constraints of this theory, containing a Schwarzschild exterior and an `empty' interior \cite{bengtsson}. Within the same theory, Varadarajan  has discovered spherically symmetric vacuum solutions with negative energy \cite{madhavan}. 

Recently, within the first order (Hilbert-Palatini) Lagrangian formulation in vacuum, generic solutions consisting of a Schwarzschild and a noninvertible tetrad phase ($\det e_\mu^I=0$) have been presented \cite{kaul}. These are associated with curvature two-form fields that are finite everywhere, in contrast to the original Schwarzschild spacetime. A class of two-sheeted geometries, which contain the Einstein-Rosen bridge \cite{einstein} as a special case, have also been found  as a set of new solutions of this theory \cite{sengupta}. These works make crucial use of a general framework for constructing  spacetime solutions whose metrics are degenerate everywhere \cite{kaul1, kaul2}. It has also been noted that geometries consisting of both the phases of the metric determinant could display intriguing causal features even in the absence of exotic matter or additional structures such as extra dimensions \cite{sandipan}.

In the next section, we elucidate on the solution based on the Taub phase which do not exhibit a Dirac string or closed timelike lines upon a smooth evolution in time. Section III contains a discussion of an emergent interpretation of charge through zero-signature geometries in vacuum gravity. Next, the programme of eliminating timelike circles is extended to the Misner spacetime, which was originally introduced as a simpler analogue of the Taub-NUT geometry. We conclude with a summary of the main results and a few worthwhile remarks.

\section{Extension of the Taub Universe}
In first-order gravity in four dimensions, the basic fields are the $SO(3,1)$ tetrad and spin-connection. These define the Hilbert-Palatini action for gravity theory in four dimensions:
\begin{equation*}
S[e,\omega]=\int d^4 x~ \epsilon^{IJKL}\epsilon^{\mu\nu\alpha\beta}e_{\mu I}e_{\nu J}R_{\alpha\beta KL}(\omega)
\end{equation*}
Note that the Lagrangian density is well-defined for both invertible and non-invertible tetrads.
Upon variation with respect to the independent fields $e_\mu^{I}$ and $\omega_{\mu}^{IJ}$, one obtains the following set of equations of motion :
	\begin{eqnarray}\label{eom}
	e_{[\mu}^{\left[ \right. I} D_{\alpha} e_{\beta]}^{J\left. \right]} =0,~	e_{[\mu}^{\left[ \right. I} R_{\alpha\beta]}^{~JK \left. \right]}&=0.
	\end{eqnarray}
The standard Einsteinian theory may be recognized as a special phase of the above, corresponding to solutions with a nonvanishing metric determinant ($\det g_{\mu\nu} \neq 0$). In general, the above set of equations exhibit vacuum solutions  which could be degenerate everywhere \cite{tseytlin,kaul1,kaul2} or almost everywhere or over a certain region \cite{kaul,einstein,sengupta,sandipan}. Moreover, such solutions may contain nonvanishing torsion associated with the degenerate phase even in the absence of any matter-coupling. Obviously, this whole set of solutions to the equations of motion are not contained within the standard Einsteinian theory. A priori, there seem to be no reason for not including the latter set of solutions within a general analysis of classical and quantum gravity \cite{tseytlin,kaul1,kaul2,horowitz}. 

With this viewpoint, we now move on to discuss the Taub-NUT spacetime in the context of first order formulation.

\subsection{The Einsteinian solution}

The Taub-NUT solution of the vacuum Einstein's equations is represented by the following metric:
\begin{equation}\label{TN}
ds^2=-f^2(r)(dt+4l \sin^2\frac{\theta}{2}d\phi)^2+\frac{dr^2}{f^2(r)}+(r^2+l^2)(d\theta^2+\sin^2\theta d\phi^2)
\end{equation}
where,
	\begin{eqnarray*}
	f^2(r)=\frac{(r-r_+)(r-r_-)}{r^2+l^2}
	\end{eqnarray*}
with $r_\pm=m\pm\sqrt{l^2+m^2}$. The two parameters $m,l$ may be interpreted as the `mass' and `magnetic charge' (or, the `dual mass'), respectively.
The ranges $r>r_+$ and $r<r_-$ define the `NUT' spacetime, whereas $r_-<r<r_+$ is the `Taub' region.  This metric  exhibits a Dirac string singularity along the half-line $\theta=\pi$. In addition, the `NUT' region contains closed timelike curves, since we have $g_{\phi\phi}<0$ for $\cos\theta<-\frac{r^2+l^2-4l^2f^2}{r^2+l^2+4l^2f^2}$ there. The zeroes of $f^2(r)$ are coordinate singularities, and there exists no curvature singularity. The spacetime, however, is known to be geodesically incomplete \cite{misner1,zimmerman,kunz}.

As originally proposed by Misner \cite{misner}, one may introduce two different time coordinates $t$ and $t'$ at $0\leq\theta<\pi$ and $0<\theta\leq\pi$ respectively, related as:
\begin{equation}\label{tt}
t'=t+4l\phi
\end{equation}
In other words, while the region $0\leq\theta<\pi$ is described by the metric \eqref{TN}, the region $0<\theta\leq\pi$ is described by a different one obtained through the above transformation:
\begin{equation}
	ds^2=-f^2(r)(dt-4l \cos^2\frac{\theta}{2}d\phi)^2+\frac{dr^2}{f^2(r)}+(r^2+l^2)(d\theta^2+\sin^2\theta d\phi^2)
\end{equation}
The atlas made up of these two charts is regular at both $\theta=0$ and $\theta=\pi$, thus eliminating the Dirac String singularity. However, the relation between $t$ and $t'$ at the overlap region  $0<\theta<\pi$ now implies that the time coordinate must have a periodicity of $8\pi l$. As a consequence, any arbitrary curve with $r=const.,~\theta=const.,~\phi=const.$ at the  NUT region represents a causality violating worldline. This feature, among others, has led to the general wisdom that the Taub-NUT spacetime cannot really be regarded as a physical solution to the Einstein's equations in vacuum \cite{misner1}.

In Einstein's theory, a Taub phase has no choice but to evolve into a NUT phase, and hence must inherit its much too well-known pathologies.
In first-order gravity, however, a Taub phase may also evolve into a degenerate phase, provided that is permissible by the boundary conditions (at the interface between the two phases). In the following, we  construct a unique degenerate phase which the Taub Universe could evolve into, without acquiring any closed timelike curve or Dirac string  anywhere. Before delving into the essential details of this solution, let us first discuss the field configuration of the Taub geometry in terms of first order variables.

According to (\ref{TN}), the Taub spacetime is defined by the region $r_-<r<r_+$ with the following metric: 
 \begin{equation}\label{T}
 ds^2=\bar{f}^2(r)(dt+4l \sin^2\frac{\theta}{2}d\phi)^2-\frac{dr^2}{\bar{f}^2(r)}+(r^2+l^2)(d\theta^2+\sin^2\theta d\phi^2),
 \end{equation}
 where $\bar{f}^2=\frac{(r_+-r)(r-r_-)}{r^2+l^2}=-f^2>0$. The coordinates $t$ and $r$ are spacelike and timelike,  respectively, in contrast to the NUT region where they interchange their roles. 
 The tetrad fields are given by:
 \begin{eqnarray*}
 e^0=\bar{f}(dt+4l \sin^2\frac{\theta}{2}d\phi),~ e^1=\frac{dr}{\bar{f}},~
 e^2=\sqrt{(r^2+l^2)}d\theta,~
 e^3=\sqrt{(r^2+l^2)}\sin\theta d\phi;
 \end{eqnarray*}
 The (torsionless) connection fields are evaluated to be:
  \begin{eqnarray*}
 \omega^{01}&=&-(\del_r\bar{f})e^0,~ \omega^{02}=\frac{l\bar{f}}{r^2+l^2}e^3,~ \omega^{03}=-\frac{l\bar{f}}{r^2+l^2}e^2,\\
 \omega^{12}&=&\frac{r\bar{f}}{r^2+l^2}e^2,~
 \omega^{23}=-\frac{l\bar{f}}{r^2+l^2}e^0-\frac{\cot\theta}{\sqrt{r^2+l^2}}e^3,~
 \omega^{31}=-\frac{r\bar{f}}{r^2+l^2}e^3
 \end{eqnarray*}
These lead to the following expressions for the $SO(3,1)$ field-strength:
\begin{align}
	R^{01}&=\left[\bar{f}\del^2_r\bar{f}+(\del_r\bar{f})^2\right]e^0\wedge e^1-\frac{2l\bar{f}}{r^2+l^2}\left[\del_r\bar{f}-\frac{r\bar{f}}{r^2+l^2}\right]e^2\wedge e^3\nonumber\\
	R^{02}&=\frac{\bar{f}}{r^2+l^2}\left[r\del_r\bar{f}+\frac{l^2\bar{f}}{r^2+l^2}\right]e^0\wedge e^2-\frac{l\bar{f}}{r^2+l^2}\left[\del_r\bar{f}-\frac{r\bar{f}}{r^2+l^2}\right]e^3\wedge e^1\nonumber\\
	R^{03}&=\frac{\bar{f}}{r^2+l^2}\left[r\del_r\bar{f}+\frac{l^2\bar{f}}{r^2+l^2}\right]e^0\wedge e^3-\frac{l\bar{f}}{r^2+l^2}\left[\del_r\bar{f}-\frac{r\bar{f}}{r^2+l^2}\right]e^1\wedge e^2 \nonumber\\
	R^{12}&=\frac{\bar{f}}{r^2+l^2}\left[r\del_r\bar{f}+\frac{l^2\bar{f}}{r^2+l^2}\right]e^1\wedge e^2-\frac{l\bar{f}}{r^2+l^2}\left[\del_r\bar{f}-\frac{r\bar{f}}{r^2+l^2}\right]e^3\wedge e^0 \nonumber\\
		R^{23}&=\frac{1}{r^2+l^2}\left[1-\frac{3\bar{f}^2 l^2-r^2 \bar{f}^2}{r^2+l^2}\right]e^2\wedge  e^3+ \frac{2l\bar{f}}{r^2+l^2}\left[\del_r\bar{f} -\frac{r\bar{f}}{r^2+l^2}\right]e^0\wedge e^1 \nonumber\\
	R^{31}&=\frac{\bar{f}}{r^2+l^2}\left[r\del_r\bar{f}+\frac{l^2\bar{f}}{r^2+l^2}\right]e^3\wedge e^1-\frac{l\bar{f}}{r^2+l^2}\left[\del_r\bar{f}-\frac{r\bar{f}}{r^2+l^2}\right]e^2\wedge e^0 
	\end{align} 
These fields represent a solution corresponding to the Einstein phase of first order gravity, since the determinant of the tetrad is nonvanishing everywhere (except at the apparent singularities of the spherical polar coordinates). 

\subsection{Solution with the zero signature phase}

In order to define a continuation of the Taub spacetime to a degenerate phase, let us introduce a new coordinate $w$ through a reparametrization of $r$:
\begin{equation}
r(w)-r_-=R(w),
\end{equation}
where  $R(w)$ is any smooth ($C^{\infty}$) function  such that\footnote{An example of such a smooth function is $R(w)=(r_+-r_-)~e^{-\frac{1}{w}}$ at $w\geq 0$.}:
\begin{equation}\label{R}
r(w\rightarrow 0)=r_-,~\left[\frac{r'(w)}{\bar{f}[r(w)]}\right](w\rightarrow 0)=0,~r(w\rightarrow \infty)\rightarrow r_+
\end{equation}
The prime introduced above implies a differentiation with respect to $w$.
Clearly, at $0< w<\infty$ the new coordinate $w$ covers the full Taub spacetime.
Next, let us consider  the  spacetime defined by the full range of the coordinates $t\in(-\infty,\infty),~w\in (-\infty,\infty),~\theta\in[0,\pi],~\phi\in[0,2\pi]$. The domains  $w>0$  and $w\leq 0$ are assumed to exhibit a Taub geometry ($\det e_\mu^I \neq 0)$ and a zero metric-determinant phase, respectively:
 \begin{eqnarray}\label{T1}
 && w>0:\nonumber\\
 && ds^2=\bar{f}^2[r(w)](dt+4l \sin^2\frac{\theta}{2}d\phi)^2-\frac{r^{'2}(w)dw^2}{\bar{f}^2[r(w)]}+\left[r^2(w)+l^2\right](d\theta^2+\sin^2\theta d\phi^2),\nonumber\\
 && w\leq 0:\nonumber\\
 &&  ds^2=0-F^2(w) dw^2 + H^2(w)(d\theta^2+\sin^2\theta d\phi^2),
 \end{eqnarray}
where $F(w)$ and $H(w)$ are smooth functions satisfying:
\begin{eqnarray}\label{F}
F(w=0)=0,~H(w=0)=\sqrt{r_-^2 +l^2}
\end{eqnarray} 
The properties given by eq. (\ref{R}) and (\ref{F}) imply that the tetrad components are smooth across the phase boundary $w=0$.
Note that the internal metric is given by $\eta_{IJ}=diag[1,-1,1,1]$ everywhere. Also, we shall use two overlapping charts as discussed earlier to cover the full spacetime. This makes the spacetime regular for all $\theta$-values and does not introduce any causality violation either at the Taub or at the degenerate phase. 

Note that the field-strength tensor components at the Taub phase are finite. At the phase boundary, their (limiting) behaviour could be found using the fact that both $e^0$ and $e^1$ vanish as $w\rightarrow 0^+$:
\begin{eqnarray}\label{bc0}
&& R^{01}\rightarrow -\frac{l(r_+-r_-)}{r_-^2+l^2}\sin\theta d\theta\wedge d\phi,~R^{02}\rightarrow 0,~R^{03}\rightarrow 0,\nonumber\\
&& R^{12}\rightarrow 0,~R^{23}\rightarrow \sin\theta d\theta\wedge d\phi,~R^{31}\rightarrow 0
\end{eqnarray}

At the zero signature phase, the tetrad fields are given by:
 \begin{eqnarray}
\hat{e}^0=0,~
\hat{e}^1=F(w)dw,~\hat{e}^2=H(w)d\theta,~\hat{e}^3=H(w)\sin\theta d\phi
\end{eqnarray}
For the connection fields, we choose the following general ansatz \cite{kaul1} which satisfies the set of connection equations of motion in (\ref{eom}) ($i\equiv (1,2,3),~a\equiv{w,\theta,\phi})$: 
\begin{eqnarray}
\hat{\omega}_a^{0i}=\lambda \hat{e}_a^i,~
\hat{\omega}_a^{ij}= \bar{\omega}_a^{ij}+\epsilon^{ijk}\hat{e}_a^l N_{kl}=\bar{\omega}_a^{ij}+K_a^{~ij},
\end{eqnarray}
where $\lambda$ is a constant, $\bar{\omega}_a^{ij}(\hat{e})$ are the torsionless connection components defined completely by the set of triads $\hat{e}_a^i$ (and their inverses) and $N_{kl}=N_{lk}$ is a matrix encoding the six independent contortion fields in $K_a^{~ij}$. Explicitly, the nonvanishing components of $\bar{\omega}_a^{ij}(\hat{e})$ and the contortion read:
\begin{eqnarray}\label{connection}
 \bar{\omega}^{12}&=&\frac{H'(w)}{F(w)H(w)}\hat{e}^2,~
 \bar{\omega}^{23}=-\frac{\cot\theta}{H(w)}\hat{e}^3,~\bar{\omega}^{31}=-\frac{H'(w)}{F(w)H(w)}\hat{e}^3
 ;\nonumber\\
 N_{kl}&=&
 \quad
\begin{bmatrix} 
\alpha(w) & \mu(w) & \delta(w) \\
\mu(w) & \beta(w) & \rho(w) \\
\delta(w) & \rho(w) & \gamma(w)
\end{bmatrix}
\quad .
 \end{eqnarray}
Here we have assumed that the contortion depends on $w$ only.

Using these, the $0i$-components of the curvature two-forms become:
\begin{align}
\hat{R}^{01}&=\lambda[\delta \hat{e}^1 \wedge \hat{e}^2+\mu \hat{e}^3 \wedge \hat{e}^1-(\beta+\gamma)\hat{e}^2 \wedge \hat{e}^3],\nonumber\\
\hat{R}^{02}&=-\lambda[\rho \hat{e}^1 \wedge \hat{e}^2+(\gamma-\alpha) \hat{e}^3 \wedge \hat{e}^1-\mu\hat{e}^2 \wedge \hat{e}^3],\nonumber\\
\hat{R}^{03}&=\lambda[(\beta-\alpha) \hat{e}^1 \wedge \hat{e}^2-\rho \hat{e}^3 \wedge \hat{e}^1+\delta \hat{e}^2 \wedge \hat{e}^3]
\end{align} 
Comparing these with the expressions in eq.(\ref{bc0}), 
we note that these components are continuous across the phase boundary $w=0~ (r=r_-
)$ provided:
\begin{eqnarray}\label{bc1}
\mu\doteq 0,~\rho\doteq 0,~\delta\doteq 0;~
\alpha\doteq \beta\doteq \gamma\doteq \frac{l (r_+-r_-)}{2\lambda (r_-^2 + l^2)^2}
\end{eqnarray} 
where the symbol $\doteq$ denotes an equality only at the phase boundary. For simplicity, we assume hereon that the three fields $\mu(w),\rho(w),\delta(w)$ vanish and that $\alpha(w)=\beta(w)=\gamma(w)$ everywhere, both set of equalities being consistent with the boundary conditions above. With this, the torsional contribution is encoded by just a single scalar $\alpha(w)$.
The remaining components of the field-strength read:
\begin{eqnarray}\label{R1}
\hat{R}^{12}&=& \left[\frac{1}{F}\left(\frac{H'}{FH}\right)'+\left(\frac{H'}{FH}\right)^2 -(\alpha^2+\lambda^2)\right] \hat{e}^1 \wedge \hat{e}^2- \left(\frac{\alpha'}{F}+\frac{2\alpha H'}{FH}\right)\hat{e}^3 \wedge \hat{e}^1\nonumber\\
\hat{R}^{23}&=& \left[\left(\frac{H'}{FH}\right)^2 +\frac{1}{H^2}+ \alpha^2-\lambda^2\right] \hat{e}^2 \wedge \hat{e}^3\nonumber\\
\hat{R}^{31}&=& \left[\frac{1}{F}\left(\frac{H'}{FH}\right)'+\left(\frac{H'}{FH}\right)^2 -(\alpha^2+\lambda^2)\right] \hat{e}^3 \wedge \hat{e}^1+ \left(\frac{\alpha'}{F}+\frac{2\alpha H'}{FH}\right)\hat{e}^1 \wedge \hat{e}^2
\end{eqnarray}
Continuity of these components at the interface $w=0$ implies a further set of boundary conditions, given by:
\begin{eqnarray}\label{bc2}
&&\left(\frac{H'}{FH}\right)'+\left[\left(\frac{H'}{FH}\right)^2 -(\alpha^2+\lambda^2)\right]F  \doteq  0,~
 \left(\frac{H'}{FH}\right)^2+ \alpha^2-\lambda^2 \doteq  0,~ \alpha'H+2\alpha H' \doteq  0
\end{eqnarray}
The last two boundary conditions above determine $\alpha(w=0)$ and $\lambda$ in terms of the two charges $l,m$. However, there is yet another equation involving these in (\ref{bc1}), which implies that the mass $(m)$ and dual mass $(l)$ are not independent of each other in this configuration.

The $SO(3,1)$-invariant fields $\hat{g}_{\mu\nu}=\hat{e}_\mu^I \hat{e}_{\nu I}$ and 
$\hat{R}_{\mu\nu\alpha\beta}=\hat{R}_{\mu\nu}^{~~IJ} \hat{e}_{\alpha I} \hat{e}_{\beta J}$ above are manifestly continuous. The only remaining set of gauge-invariant fields are the contortion $K_{\mu\nu\alpha}=K_{\mu}^{~IJ}\hat{e}_{\nu I} \hat{e}_{\alpha J}$, whose nonvanishing components given by:
\begin{eqnarray}
K_{\phi w\theta}=-\alpha FH^2\sin\theta=K_{\theta\phi w}=-K_{w\theta\phi}.
\end{eqnarray}
Evidently, $K_{\mu\nu\alpha}$ vanish at the phase boundary, as is required by continuity with the torsionless Taub phase across $w=0$.  The crucial role of torsion at the degenerate phase is evident, since it is not possible to achieve the continuity of field-strength components without its presence.

As mentioned earlier, this configuration based on the general ansatz (\ref{connection}) for the connection fields satisfies the first set of the equations of motion in (\ref{eom}) by construction.
The remaining set is also satisfied provided the contortion field $\alpha(w)$ is constrained as:
\begin{eqnarray}\label{mc}
\left[\alpha^2-3\left(\frac{H'}{FH}\right)^2 -\frac{1}{H^2}+ 3\lambda^2\right]F-2\left(\frac{H'}{FH}\right)'=0
\end{eqnarray}
Since this is the only equation among the three unknowns $F(w),H(w)$ and $\alpha(w)$, it can always be solved for any of them by choosing the other two appropriately. 
A maximal extension of the zero-signature phase is obtained by imposing an additional condition on $F(w)$:
\begin{eqnarray}
|\int_0^{-\infty}dw~F(w)|=\infty,
\end{eqnarray}
which implies that the boundary $w\rightarrow -\infty$ is approached at an infinite value of the affine parameter.

With this, the full spacetime geometry is completely characterized by the fields $(e_{\mu}^I,\omega_\mu^{IJ})$ defining  a Taub phase at $ w>0$ and $(\hat{e}_{\mu}^I,\hat{\omega}_\mu^{IJ})$ associated with a non-invertible metric phase at  $w\leq 0$.
This solution is smooth everywhere. It could be interpreted as a union of two `Universes', which are causally disconnected in the sense of geodesics since the phase boundary $w=0$ separating them is time non-orientable. In terms of the coordinate time $w$ going from $-\infty$ to $\infty$ ($\infty$ to $-\infty$), the spacetime evolves from a Lorentzian phase of three metrical dimensions to a Taub Universe which is nucleated at $w=0$. The boundary $w\rightarrow \infty$ signifies the end (beginning) of time, where the spacetime approaches a spatial two-sphere.


To emphasize, the Taub phase does not evolve into a NUT region in this continuation. This is so because the pair of boundaries of the Taub phase here ($w=0$ and $w\rightarrow \infty$) are inequivalent to the boundaries of the original Taub universe, the two pairs being related through a vanishing Jacobian. 

Importantly, the full spacetime solution does not contain any timelike circle. Neither does it carry a Dirac string, which has already been eliminated at the Taub phase by using two overlapping coordinate charts \cite{misner} and does not reappear at the zero-signature phase. 

Another crucial aspect is the finiteness of the associated field strength components (everywhere). At the degenerate phase, the effective three-geometry may also be thought to be described by a set of nondegenerate triad fields $\tilde{e}_a^i\equiv\hat{e}_a^i$ and the torsionful connection fields $\hat{\omega}_a^{ij}$. Within such a description, it is possible to construct scalar polynomials based on the effective three-curvature tensor $\hat{R}_{ab}^{~cd}=\hat{R}_{ab}^{~ij}\tilde{e}^c_i \tilde{e}^d_j$, where $\tilde{e}^a_i$ are the inverses of the triads $\tilde{e}_a^i$. All these scalars are finite, as follows from eq.(\ref{R1}). In this sense, the geometry at the degenerate phase does not exhibit any curvature divergence.

\subsection{Uniqueness Of The Degenerate Extension}
One wonders if there could be possible  degenerate extensions of the Taub Universe other than the one developed here in first order gravity, where the degenerate phases could be defined through either $g_{ww}=0$ or $g_{\theta\theta}=0$ or $g_{\phi\phi}=0$. In the first case, there exists no time coordinate which can go over the full range, and neither does it allow continuity of the field-strength across the phase boundary in general. A degenerate phase defined through a null angular coordinate $\theta$ or $\phi$, on the other hand, does not lead to an extension of the original Taub universe to a larger space without closed timelike curves. Thus, in a meaningful sense, the extension of the Taub geometry developed here could be thought to be unique, the defining properties being the following:

(i) The resulting spacetime exhibits neither a Dirac string singularity nor any closed timelike curve anywhere;

(ii) There exists a local timelike coordinate almost everywhere (with the possible exception at the asymptotic and degenerate phase boundaries), which could be extended to infinite values in both directions;

(iii) The affine parameter (for any arbitrary finite range) within the degenerate region is a monotonic function of the coordinate time;

(iv) The $SO(3,1)$ invariant fields ($g_{\mu\nu}$, $K_{\mu\nu\alpha}$ and $R_{\mu\nu\alpha\beta}$) are smooth;

(v) The curvature two-form fields are finite everywhere in spacetime.


\section{Origin of Charge (without matter)}

The  NUT solution was originally found as an empty space generalization of the Schwarzschild geometry, containing an extra parameter in the form of a dual mass \cite{nut}. However, unlike the full Schwarzschild spacetime, the full Taub-NUT geometry possesses no curvature singularity at $r=0$. The spacetime may be extended freely through this point. 
Hence, within Misner's interpretation of this solution without a Dirac string but with a periodic time, it is not clear  what is the location of the source, whose gravitational and magnetic charges are $m$ and $l$, respectively. 

However, no such interpretational difficulty  appears here. Whereas the Taub phase $w>0$ has a vanishing  contortion and the Bianchi identity is satisfied there, the noninvertible phase at $w<0$ sources (geometric) torsion and violates the Bianchi identity:
\begin{eqnarray*}
\hat{R}_{[w\theta\phi]}^{~~~~~0}=-2\lambda \alpha FH^2\sin\theta~.
\end{eqnarray*}
This violation is precisely the characteristic of a monopole, leading to
 a natural interpretation of the zero-signature geometry itself as the source of the charge $l(m)$ (in the absence of any Dirac string singularity). This is quite similar in spirit, but somewhat different in essence to the remarkable idea that couplings associated with electromagnetic interactions may originate purely due to nontrivial geometry or topology (through wormholes or geons or nonorientability of space), as pioneered by Wheeler, Misner and explored by others \cite {wheeler,wheeler1,sorkin}. 

It is also worthwhile to observe another feature, namely, a potential topological interpretation of the magnetic charge  in terms of a torsional three-integral related to the Nieh-Yan invariant \cite{nieh} in gravity theory. To this end, let us note that the extended vacuum solution is associated with a nontrivial value for the Nieh-Yan  three-form at the degenerate phase, whose integral reads:
\begin{eqnarray}\label{NY}
C_{NY}=\int d^3 x~\hat{e}^I \wedge \hat{T}_I=2 \int \alpha(w)~ \hat{e}^1 \wedge \hat{e}^2 \wedge \hat{e}^3=8\pi \int _{-\infty}^{0} dw~ \alpha(w) F(w) H^2(w)
\end{eqnarray}
Here we have used the explicit expression for the torsion $\hat{T}^I$ at the degenerate phase, given by:
\begin{eqnarray}
\hat{T}^0=0=\hat{T}^2=\hat{T}^3,~\hat{T}^1=-2\alpha(w)~ \hat{e}^2 \wedge \hat{e}^3.
\end{eqnarray} 
Note that the Nieh-Yan integral (\ref{NY}) is manifestly a function of $l(m)$ only. 
However, this torsional interpretation of the magnetic charge acquires a precise topological meaning only under a suitable compactification of the manifold through Euclidean methods or otherwise.

The above fact also implies that classically, the monopole charge is not quantized in this extended spacetime in general, provided the (gauge-invariant) field components are assumed to be continuous as here. 
This absence of charge quantization  is indeed consistent with expectations, since the time coordinate is not periodic in our construction unlike in Misner's acausal interpretation of the Taub-NUT solution \cite{dowker}.

\section{Misner Geometry without timelike loops}
There does exist a simple two-dimensional analogue of the Taub-NUT vacuum solution, which exhibits quite similar causal pathologies:
\begin{eqnarray}\label{misner1}
ds^2 =-\frac{dt^2}{t}+t d\phi^2
\end{eqnarray}
where $-\infty<t<\infty$ and $\phi\in [0,2\pi]$ is periodic.
Through a coordinate transformation $\phi=\psi+ln|t|$, this may be brought to the form in which it was originally introduced by Misner \cite{misner1}:
\begin{eqnarray*}\label{misner2}
ds^2 =2dt d\psi+t d\psi^2
\end{eqnarray*}
This  space is a flat solution of Einstein's theory. As is evident, the coordinate $t$ is timelike for $t>0$ and spacelike for $t<0$.  Any arbitrary $t=const<0$ slice traces out a closed timelike curve. The closed null curve at $t=0$ acts as the chronology horizon separating the causal and acausal parts of the spacetime. There also exists a closed null geodesic (although of zero proper length). In terms of an affine parameter $\lambda$, this reads:
\begin{eqnarray*}\label{null}
t=0,~\psi=-ln \lambda^2+const.
\end{eqnarray*} 

We shall now discuss the essential details of the solution of first order gravity containing a Misner phase, which do not contain any closed timelike or null worldline. 
Since we would prefer to keep our discussion in a four dimensional setting, we shall consider the four dimensional generalization of Misner's metric hereon:
\begin{eqnarray}\label{misner}
ds^2 =-\frac{dt^2}{t}+t d\phi^2+dy^2+dz^2
\end{eqnarray}
 

 
\subsection*{A unique degenerate extension of Misner phase}
We consider a smooth spacetime defined through the following metric, exhibiting both the phases of first order gravity at different domains:
\begin{eqnarray}
ds^2&=&-\frac{f^{'2}(t)}{f(t)}dt^2+f(t)d\phi^2+dy^2+dz^2 \mathrm{~~~at~}t>0,\nonumber\\
&=&-F^2(t)dt^2+0+dy^2+dz^2 \mathrm{~~~~at~}t\leq 0,
\end{eqnarray}
 where $-\infty<t<\infty,~0\leq \phi\leq 2\pi,~-\infty<y<\infty,~-\infty<z<\infty$. The internal metric is defined as $\eta_{IJ}=diag[-1,1,1,1]$ everywhere.
 The smooth functions $f(t)$ and $F(t)$ satisfy the following properties, ensuring continuity of the basic fields:
 \begin{eqnarray}
 && f(t)\rightarrow 0,~\frac{f^{'2}(t)}{f(t)}\rightarrow 0 ~~as~ t\rightarrow 0~;\nonumber\\
 && F(t=0)= 0 ~.
 \end{eqnarray}
 These functions also define the affine parameter $\lambda(t)$ as (for any stationary observer):
\begin{eqnarray}
\lambda(t)&=&\int_0^t dt~ \frac{f'(t)}{\sqrt{f(t)}}~~(t>0),\nonumber\\
&=&\int_0^t dt~F(t)~~(t\leq 0)
\end{eqnarray} 
 Hence, $f(t)$ at $t>0$ and $F(t)$ at $t<0$ must be such that the affine parameter is a monotonic function of the time coordinate over its full range (ruling out the possibility of a `time travel' in proper time, a phenomenon discussed in ref.\cite{sandipan}). Moreover, we make the degenerate extension maximal by demanding that the asymptotic boundaries $t\rightarrow\pm\infty$ be located at an infinite distance in the affine parameter.\footnote{An explicit example of the set of smooth functions $\left(f(t),F(t)\right)$ satisfying all the properties being discussed here is given by: $f(t)=\frac{t^2}{4}e^{-\frac{2}{t}},~F(t)=\left(1-\frac{1}{t}\right)e^{\frac{1}{t}}$.}

 Note that at $t>0$ (but not at $t\leq 0$), the metric is equivalent to the Misner spacetime (\ref{misner}) upto a coordinate redefinition $f(t)\rightarrow t$. Our task is to construct a vacuum solution of the first order field equations (\ref{eom}) based on this metric. This requires a complete specification of the spin-connection fields, which may be assumed to have vanishing torsion everywhere.
  
  The nonvanishing components of the associated fields in the invertible metric phase at $t>0$ are given by:
 \begin{eqnarray}
  && e^0=\frac{f'(t)}{\sqrt{f(t)}}dt,~e^1=\sqrt{f(t)}d\phi,~e^2=dy,~e^3=dz;\nonumber\\
 && \omega^{01}=\frac{1}{2}d\phi.
 \end{eqnarray}
 The zero-determinant phase at $t\leq 0$ is defined by the following configuration:
\begin{eqnarray}
&& \hat{e}^0=F(t)\cosh\left(\frac{\phi}{2}\right)dt,~\hat{e}^1=-F(t)\sinh\left(\frac{\phi}{2}\right)dt,~\hat{e}^2=dy,~\hat{e}^3=dz;\nonumber\\
&&\hat{\omega}^{01}=\frac{1}{2}d\phi.
\end{eqnarray}
These fields are manifestly continuous across the phase boundary. Since the torsion as well as the curvature two-form fields vanish everywhere, both the set of first order field equations (\ref{eom}) are trivially satisfied by this configuration. This smooth geometry is a unique (maximal) extension of the Misner spacetime in the same sense as elaborated in the previous section.

Let us now illustrate that there are no closed timelike solutions to the geodesic equations in this geometry. At $t>0$, these equations lead to the following constants of motion:
 \begin{eqnarray}
 \frac{f^{'2}(t)}{f(t)}\dot{t}^2-f(t)\dot{\phi}^2-\dot{y}^2-\dot{z}^2=\epsilon,~f\dot{\phi}=p_\phi,~\dot{y}=p_y,~\dot{z}=p_z,
 \end{eqnarray}
 where $\epsilon=0,+1,-1$ characterize the null, timelike and spacelike geodesics, respectively. The solution for the affine parameter is:
 \begin{eqnarray}
 \lambda=\pm \int dt~\frac{f'(t)}{\sqrt{(\epsilon+p_y^2+p_z^2)f(t)+p_\phi^2}}=\pm\frac{2}{\sqrt{\epsilon+p_y^2+p_z^2}}\sqrt{f(t)+\frac{p_\phi^2}{\epsilon+p_y^2+p_z^2}}
 \end{eqnarray}
  At the noninvertible phase $t\leq 0$, the geodesic equations corresponding to the $t,y$ and $z$ coordinates read:
 \begin{eqnarray}
 F^2(t)\dot{t}^2-\dot{y}^2-\dot{z}^2=\epsilon,~\dot{y}=p_y,~\dot{z}=p_z, 
 \end{eqnarray}
 while the equation for the null direction $\phi$ becomes redundant. This is expected since the $\phi$ coordinate in this region has no physical evolution. The above set of equations are solved as:
 \begin{eqnarray}
 \lambda=\pm\frac{1}{\sqrt{\epsilon+p_y^2+p_z^2}}\int dt~F(t)
 \end{eqnarray}
 These solutions completely specify the geodesics over the full spacetime, which contain no closed timelike trajectories. There is no closed null geodesic also, since there is no compact direction to source such trajectories at the phase boundary ($t=0$) which now replaces the chronology horizon. 
 
Note that unlike the Taub-NUT case, the presence of torsion in the degenerate geometry is not essential in order to satisfy the continuity of field strength components. This can be attributed to the fact that the original Misner space does not contain any analogue of magnetic charge (Dirac string).

 \section{Conclusions}
 
 The Taub-NUT vacuum solution of Einstein's equations, as it is, does not admit a completely reasonable physical  interpretation. Even though it contains the Taub Universe which could have a realization as a time-dependent cosmology with non-isotropic spatial sections, this phase has a finite lifetime. Beyond this span, it naturally evolves to the NUT geometry, which exhibits robust singularities in the form of a Dirac string along a half-axis as well as closed timelike lines over a region. In this work, we have demonstrated that both of these pathologies can be eliminated within the first order formulation of gravity provided the Taub geometry
 coexists with a zero-determinant metric phase. This defines a smooth spacetime that satisfies the vacuum field equations everywhere. The special way in which this coexistence has to happen makes this extension of the Taub universe unique in a precise sense.  To pursue this idea to a completion in the context of vacuum solutions with similar causal singularities, we have also presented a continuation of the Misner phase, which in its Einsteinian description is known to exhibit closed timelike curves. 
 
Although the differences between the original Taub-NUT or Misner spacetime and their generic counterparts  in first-order formulation are really the reasons why the latter are being discussed here, there does exist important similarities. In particular, both are geodesically incomplete, but nevertheless are free of any divergence in the curvature two-form fields.  
One wonders whether there could be a sense in which the spacetimes constructed here could be seen as the more probable configurations compared to the Einsteinian ones in the quantum theory.

Finally, in our analysis based on the Taub phase, zero signature geometries emerge as a possible origin
 of couplings (charges) apparently associated with gravitational or electromagnetic interactions. This emergent picture does not require a violation of the vacuum field equations anywhere, unlike the Wheeler-Misner programme based on geons and wormholes, for instance. It is also worth emphasizing that the violation of Bianchi identity, as is the characteristic signature of a magnetic charge (with or without matter), does not imply a loss of smoothness of the (gauge-invariant) fields here. This may be contrasted with the approach where a violation of such symmetries is exhibited through non-smooth metrics \cite{kuntz}. In addition, the magnetic charge here is explicitly given in terms of a torsional three-integral related to the Nieh-Yan (topological) invariant.
This connection, which have not really been explored here, may have topological underpinnings and deserves further thoughts.

\vspace{.2cm}

\acknowledgments 
S.S. is indebted to M. Varadarajan for instructive discussions on degenerate triad solutions in the $SU(2)$ Hamiltonian framework, to M. Campiglia for helpful criticisms of a preliminary draft and also to G. Date and R. Kaul for commenting on it.
This author's work is supported (in part) by the ECR/2016/000027 grant, SERB, DST, Government of India. S.G. gratefully acknowledges the support of a DST-Inspire fellowship.

\end{document}